\documentclass[superscriptaddress,nofootinbib,notitlepage]{revtex4-1}

\usepackage{amsmath,amssymb}
\usepackage{graphicx}

\begin{document}

\title{$\Delta(27)$ family symmetry and neutrino mixing}
\author{Ivo de Medeiros Varzielas}
\email{ivo.de@soton.ac.uk}
\affiliation{{\small School of Physics and Astronomy, University of Southampton,}\\
Southampton, SO17 1BJ, U.K.}


\begin{abstract}
The observed neutrino mixing, having a near maximal atmospheric neutrino mixing angle and a large solar mixing angle, is close to tri-bi-maximal.
This structure may be related to the existence of a discrete non-Abelian family symmetry.

In this paper the family symmetry is the non-Abelian discrete group $\Delta(27)$, a subgroup of $SU(3)$ with triplet and anti-triplet representations. 
Different frameworks are constructed in which the mixing follows from combining fermion mass terms with the vacuum structure enforced by the discrete symmetry.
Mass terms for the fermions originate from familon triplets, anti-triplets or both.
Vacuum alignment for the family symmetry breaking familons follows from simple invariants.
\end{abstract}

\maketitle

\section{Introduction}

The observed neutrino oscillation parameters \cite{Capozzi:2013csa, Forero:2014bxa,Gonzalez-Garcia:2014bfa} are consistent with a nearly tri-bi-maximal (TBM) structure
\begin{equation}
V_{TBM}= \left[ 
\begin{array}{ccc}
-\sqrt{\frac{4}{6}} & \sqrt{\frac{1}{3}} & 0 \\ 
\sqrt{\frac{1}{6}} & \sqrt{\frac{1}{3}} & \sqrt{\frac{1}{2}} \\ 
\sqrt{\frac{1}{6}} & \sqrt{\frac{1}{3}} & -\sqrt{\frac{1}{2}}%
\end{array}%
\right]  \label{eq:TBM}
\end{equation}
It has been observed that this simple form might be a hint of an underlying family symmetry.
The observation of non-zero $\theta_{13}$ of comparable magnitude to the Cabibbo angle excludes exact TBM, motivating the search for frameworks where $\theta_{13}$ is naturally of the observed magnitude and the remaining angles are close to their TBM values.

The discrete non-Abelian group $\Delta(27)$ is an interesting candidate family symmetry for such frameworks.
It is a subgroup of $SU(3)$ of order 27 with a triplet and anti-triplet representation. After $T_7$ (order 21) it is the smallest group with this appealing feature, and the smallest in the series of groups $\Delta(3n^2)$, which contains $C_3$, $A_4$ and $\Delta(27)$ for $n=1,2,3$ respectively.

In terms of use in particle physics $\Delta(27)$, has an interesting history. It was used over 50 years ago in hadron physics in \cite{Fairbairn:1964sga}, and over 30 years ago to obtain spontaneous geometrical CP violation in \cite{Branco:1983tn}, where the vacuum expectation values (VEVs) display a CP violating phase that is related to the group structure and independent of arbitrary phases in the Lagrangian. Interest in discrete groups was rekindled with neutrino oscillations providing additional pieces of the flavour puzzle, and $\Delta(27)$ was rediscovered and used to describe fermion masses and mixing, first by \cite{deMedeirosVarzielas:2006fc}, followed closely by \cite{Ma:2006ip} - see also \cite{deMedeirosVarzielas:2006bi,Ma:2007wu,Varzielas:2008kr,Grimus:2008tt,Bazzocchi:2009qg,Howl:2009ds}.
Although the group details were included in \cite{Fairbairn:1964sga}, with the renewed interest in discrete groups the group theoretical details were presented conveniently and in great detail in more recent papers \cite{Luhn:2007uq, Ludl:2009ft, Ishimori:2010au}.

$\Delta(27)$ (together with $\Delta(54)$, the $n=3$ group in the series $\Delta(6n^2)$ \cite{Escobar:2008vc}) has received additional interest due to its CP properties, that had been first explored in \cite{Branco:1983tn}. A more systematic analysis was started by \cite{deMedeirosVarzielas:2011zw}, and followed by a series of papers aiming to further clarify or implement spontaneous geometrical CP violation
\cite{Varzielas:2012nn,Varzielas:2012pd,Bhattacharyya:2012pi,Ivanov:2013nla,Varzielas:2013zbp,Ma:2013xqa,Varzielas:2013sla,Varzielas:2013eta}.
Additionally, a model based on $\Delta(27)$ obtaining maximal CP violation in the lepton sector was proposed in \cite{Ferreira:2012ri}.
Still with respect to the CP properties of $\Delta(27)$, \cite{Holthausen:2012dk, Chen:2014tpa, Fallbacher:2015rea} explored the important role of the group automorphisms, while 
 \cite{Nishi:2013jqa} listed all CP transformations consistent with $\Delta(27)$ triplets.

Recently, \cite{Branco:2015hea} proposed to use an invariant approach to CP in the presence of family symmetries. Using it to analyse a $\Delta(27)$ model revealed the group featured also geometrical CP violation without spontaneous symmetry breaking - explicit geometrical CP violation. A more complete analysis of $\Delta(27)$ Lagrangians with the invariant approach followed in \cite{Varzielas:2015fxa,Branco:2015gna}.

While the CP properties of $\Delta(27)$ have clearly been the focus of recent attention, the role of $\Delta(27)$ in explaining fermion mixing deserves further interest, particularly as it is a small group with the potential to lead to models with viable leptonic mixing.
While there have been a few recent works based on $\Delta(27)$, the models with a dark matter candidate presented in \cite{Ma:2014eka} are simply based on \cite{Ma:2006ip, Ma:2007wu}, whereas the novel proposals \cite{Aranda:2013gga,Harrison:2014jqa,Abbas:2014ewa} have multiple triplet familon VEVs and no clear mechanism for keeping the different VEV directions from perturbing each other. Arguably, there have been no simple models demonstrating clearly how $\Delta(27)$ would generate both the special VEVs and the associated leptonic mixing.

Although \cite{deMedeirosVarzielas:2006bi} showed 9 years ago how to obtain relevant familon VEVs in supersymmetric (SUSY) frameworks with $\Delta(27)$ and similar groups, the mechanism referred to as D-term alignment hasn't been widely adopted, possibly due to its dependence on soft SUSY breaking terms. D-term alignment was used also in the $\Delta(27)$ models of \cite{Bazzocchi:2009qg, Howl:2009ds} as well as in the $T_7$ model of \cite{Luhn:2007sy} and the $A_4$ model of \cite{Varzielas:2008jm}. Some strategies for making D-term alignment less dependent on the relative magnitudes of familon VEVs were discussed in \cite{deMedeirosVarzielas:2011wx} and \cite{Varzielas:2012ss}.

The goal of this paper is to construct simple $\Delta(27)$ frameworks in analogy with \cite{Altarelli:2005yx, deMedeirosVarzielas:2005qg, Altarelli:2009kr, Varzielas:2010mp,Varzielas:2012ai}, which separately align multiple familons in SUSY frameworks. Simple framework are well known for $A_4$ \cite{Altarelli:2005yx, deMedeirosVarzielas:2005qg, Altarelli:2009kr, Varzielas:2010mp,Varzielas:2012ai} and $S_4$ \cite{Varzielas:2012pa} (see also \cite{Bazzocchi:2009pv,Dutta:2009bj, BhupalDev:2012nm} for non-renormalisable alignment terms in $S_4$).
$\Delta(27)$ is very distinct from the well known cases both in the alignment of the familons and the construction of fermion invariants, due to its main advantages over those groups: the large number of singlet representations which is partly related to its interesting CP properties, and its triplet and anti-triplet representations which make the group particularly suited for grand unification. With separate triplet and anti-triplet, groups like $T_7$ \cite{Luhn:2007sy}, \cite{Hagedorn:2008bc, Vien:2014gza, Bonilla:2014xla, Hernandez:2015cra}, $\Delta(27)$ and larger $\Delta(3n^2)$ or $\Delta(6n^2)$ groups \cite{Varzielas:2012ss}, \cite{Ishimori:2008uc,Ishimori:2009ew,Escobar:2011mq, Ding:2012xx, King:2012in, Lam:2013ng} forbid the triplet-triplet invariant and naturally avoid a problematic contribution to triplet mass terms (proportional to the identity matrix).

In Section \ref{sec:VEVs} the simplest options to align familon VEVs with F-terms are shown, significantly improving the basic ideas proposed in \cite{deMedeirosVarzielas:2005qg}, and providing an alternative to the D-term alignment \cite{deMedeirosVarzielas:2006bi, deMedeirosVarzielas:2011wx, Varzielas:2012ss}. One of the possibilities found allows to obtain in SUSY frameworks the $(\omega,1,1)$ class of VEV involved in spontaneous geometrical CP violation \cite{Branco:1983tn,deMedeirosVarzielas:2011zw}.
Section \ref{sec:masses} features several frameworks that couple the fermions to the familons. To complete these frameworks, a method of safeguarding against terms that would invalidate them is required. 
An auxiliary symmetry protecting a minimal framework is presented in Section \ref{AS}, in order to show a specific example of a complete framework with $\Delta(27)$ family symmetry and neutrino mixing.
Finally in Section \ref{sec:Conclusion} the conclusions are presented.

\section{$\Delta(27)$ family symmetry}

A complete family symmetry model is usually constituted by alignment of familon VEVs (Section \ref{sec:VEVs}), the coupling of the familons to fermions (Section \ref{sec:masses}), and methods to eliminate terms that would invalidate the model (Section \ref{AS}).
Appendix \ref{D27} contains the relevant details about $\Delta(27)$, namely product rules for the representations in the notation used throughout the paper to build invariants.

\subsection{Aligning the familons \label{sec:VEVs}}

An important advantage of building family symmetry models in SUSY frameworks is the holomorphic superpotential. This facilitates the separation of distinct familons as noted in \cite{Altarelli:2005yx, deMedeirosVarzielas:2005qg}, through F-term alignment involving alignment superfields (often referred to as driving fields). Although D-term alignment can be more minimal in the sense that it dispenses the introduction of alignment fields \cite{deMedeirosVarzielas:2006bi, deMedeirosVarzielas:2011wx, Varzielas:2012ss}, F-term alignment can often proceed through simpler invariants.

Although there are some interesting non-renormalisable alignment terms including the one introduced in \cite{deMedeirosVarzielas:2005qg}, here the focus is exclusively on renormalisable F-term options. These are preferred in particular because in standard UV completions of non-renormalisable alignment terms some messengers act as additional alignment fields and usually spoil the desired alignment, as pointed out in \cite{Varzielas:2012ai}.

The simplest possibilities for aligning the familon VEVs used throughout the paper are described in this Section: triplet alignment fields $\varphi$ and specific singlet familons align anti-triplet familons $\bar{\phi}$, similarly anti-triplet alignment fields $\bar{\varphi}$ align triplet familons $\phi$, and singlet alignment fields $\varsigma$ lead to relative alignment between triplet and anti-triplet familons.
Appendix \ref{align} discusses these and other options in more detail, as well as their applicability to other discrete non-Abelian groups that have similar product rules and representations.

Triplet alignment field $\varphi_{1}$ combined with $1_{i0}$ singlet familons leads to VEVs with two zeros:
\begin{align}
a_{00} [\varphi_{1} \bar{\phi}_{1}]_{00} \sigma_{00} + a_{10} [\varphi_{1} \bar{\phi}_{1}]_{20} \sigma_{10}
\label{a1_align}
\end{align}
the F-term with respect to the $\varphi_{1_i}$ components giving conditions on the components $\bar{\phi}_1^i$ of the familon $\bar{\phi}_1$
\begin{align}
&a_{00} \bar{\phi}_{1}^1 \sigma_{00} + a_{10} \bar{\phi}_{1}^1 \sigma_{10} = 0 \\
&a_{00} \bar{\phi}_{1}^2 \sigma_{00} + a_{10} \omega \bar{\phi}_{1}^2 \sigma_{10} = 0 \\
&a_{00} \bar{\phi}_{1}^3 \sigma_{00} + a_{10} \omega^2 \bar{\phi}_{1}^3 \sigma_{10} = 0
\end{align}
which force two entries of the VEV to vanish, depending on the singlet familon VEVs one solution is
\begin{align}
\langle \bar{\phi}_{1} \rangle = (\bar{a}_1,0,0) \,, \quad  \frac{\langle \sigma_{10} \rangle}{\langle \sigma_{00} \rangle} = -\frac{a_{00}}{a_{10}}
\end{align}

Triplet alignment field $\varphi_{123}$ combined with $1_{0i}$ singlet familons leads instead to VEVs with equal entries:
\begin{align}
c_{00} [\varphi_{123} \bar{\phi}_{123}]_{00} \sigma_{00} + c_{01} [\varphi_{123} \bar{\phi}_{123}]_{02} \sigma_{01}
\label{c_align}
\end{align}
the F-term with respect to the $\varphi_{123_i}$ components giving conditions on the components $\bar{\phi}_{123}^i$ of the familon $\bar{\phi}_{123}$
\begin{align}
&c_{00} \bar{\phi}_{123}^1 \sigma_{00} + c_{01} \bar{\phi}_{123}^2 \sigma_{01} = 0 \\
&c_{00} \bar{\phi}_{123}^2 \sigma_{00} + c_{01} \bar{\phi}_{123}^3 \sigma_{01} = 0 \\
&c_{00} \bar{\phi}_{123}^3 \sigma_{00} + c_{01} \bar{\phi}_{123}^1 \sigma_{01} = 0
\end{align}
one solution forces the 3 entries of the VEV to be equal
\begin{align}
\langle \bar{\phi}_{123} \rangle = (\bar{c},\bar{c},\bar{c}) \,, \quad  \frac{\langle \sigma_{01} \rangle}{\langle \sigma_{00} \rangle} = -\frac{c_{00}}{c_{01}}
\end{align}

Although it isn't used further in this paper, it is relevant to note that using singlet familons $1_{ij}$ with $i,j \neq 0$ in this type of term enables the alignment of directions such as $(\omega,1,1)$ in a SUSY framework. This class of VEV is relevant in that it leads to spontaneous geometrical CP violation \cite{Branco:1983tn,deMedeirosVarzielas:2011zw}. While the VEV has clearly been obtained in non-SUSY frameworks, a way to align this direction in a SUSY framework had not been presented so far.

If aligning a single familon direction, the alignment field and familons can be neutral under additional symmetries and the $\sigma_{00}$ is superfluous. In order to have both alignments simultaneously requires alignment fields and familons separated by some mechanism, usually auxiliary symmetries - see Section \ref{AS} for a specific example.

Analogous terms relying on anti-triplet alignment fields can align triplet familons.

Assuming $\langle \phi_{1} \rangle$, $\langle \phi_{123} \rangle$ triplet VEVs have been aligned, singlet alignment fields lead to the orthogonal anti-triplet VEV 
\begin{align}
\varsigma_{02} [\phi_{123} \bar{\phi}_{23}]_{01}
\end{align}
of which one solution is
\begin{align}
\langle \bar{\phi}_{23} \rangle = (0,-\bar{b},\bar{b})
\end{align}
where a trivial singlet alignment field $\varsigma_{00}$ imposing orthogonality with $\langle \phi_{1} \rangle$ can guarantee $\langle \bar{\phi}_{23}^1 \rangle=0$.\footnote{
This $\langle \bar{\phi}_{23} \rangle$ can play a relevant role in explaining lepton non-universality in models with leptoquarks, see \cite{Varzielas:2015iva}.}
The representation of $\varsigma_{02}$ can be any of the $1_{0i}$ due to the equal components of $\langle \phi_{123} \rangle$.
Note that the orthogonality is between triplets and anti-triplet (or vice-versa).

Another relevant VEV direction is
\begin{align}
\langle \bar{\phi}_{3} \rangle = (0,0,\bar{a}_3)
\end{align} 
which can be obtained as another solution of eq.(\ref{a1_align}) or generalising it to use two non-trivial singlets. If the anti-triplet familon is accompanied by a triplet familon, the direction with two zeros for one of the familons can be obtained by combining 3 singlet alignment fields, as discussed in more detail in Appendix \ref{align} see e.g. eqs.(\ref{alt1},\ref{alt2},\ref{alt3}).

If required for contractions with a triplet into singlets, $\bar{\phi}_{1}$ and $\bar{\phi}_{3}$ can be used interchangeably if used in conjunction with $1_{0i}$ singlets e.g. $[L \bar{\phi}_{1}]_{00} \sigma_{00} \sim [L \bar{\phi}_{3}]_{01} \sigma_{02}$ as both isolate $L_1$, the first component of triplet $L$. In this sense, having both VEV directions is often not necessary in $\Delta(27)$ frameworks, as demonstrated by the specific example in Section \ref{AS}, featuring a minimal alignment superpotential $S_V$.

\subsection{Building the fermion mass terms \label{sec:masses}}

The Standard Model (SM) $SU(2)_{L}$ doublet leptons are contained in $L$, which is assigned as a triplet of $\Delta(27)$ and $e^c$, $\mu^c$, $\tau^c$ can be singlets. $H_u$, $H_d$ are the two $SU(2)_{L}$ doublets required in SUSY frameworks and are trivial singlets of $\Delta(27)$.
The charged lepton mass matrix can be made diagonal in the basis where the familon VEVs are presented, referred to henceforth as the familon basis.
One option is using an anti-triplet familon $\langle \bar{\phi}_{3} \rangle=(0,0,\bar{a}_3)$ together with trivial singlets $e^c$, $\mu^c$, $\tau^c$ and a non-trivial $1_{01}$ singlet familon $\sigma_{01}$ 
\begin{align}
S_C =  H_d \left( \frac{1}{M} [L \bar{\phi}_3]_{00} \tau^c+ \frac{1}{M^2} [L \bar{\phi}_3]_{02} \sigma_{01} \mu^c + \frac{1}{M^3} [L \bar{\phi}_3]_{01} \sigma_{01}^2 e^c \right)
\label{cl}
\end{align}
where coefficients are implicit and $M$ represents the masses of messenger fields which would be specified in a specific UV completion of the family symmetry model \cite{Varzielas:2010mp, Varzielas:2012ai}. In the remainder of the paper $M$ is omitted in the non-renormalizable superpotential terms.
If $\sigma_{01}$ in eq.(\ref{cl}) is not neutral under an additional auxiliary symmetry (as discussed in more detail in Section \ref{AS}), the different powers of $\sigma_{01}$ can match different charges of $e^c$, $\mu^c$, $\tau^c$ under the auxiliary symmetry in a version of the Froggatt-Nielsen mechanism \cite{Froggatt:1978nt}. In this way $\frac{\langle \sigma_{01} \rangle}{M}$ explains the hierarchy of the charged lepton masses, similarly to \cite{Altarelli:2009kr}. Another option is to have non-trivial singlets $e^c$, $\mu^c$, $\tau^c$ similarly to \cite{Altarelli:2005yx}, and either add a separate Froggatt-Nielsen symmetry or leave the hierarchy unexplained as in the SM.

With the charged leptons diagonal in the familon basis, all the leptonic mixing is present in the neutrino sector. It will be determined in particular by a familon VEV in the $(1,1,1)$ direction, which is an ingredient in all of the frameworks discussed here.

The effective neutrino terms can be obtained through a seesaw mechanism, such as type I seesaw. In such a case one can include Dirac terms featuring explicit geometrical CP violation \cite{Branco:2015hea,Varzielas:2015fxa,Branco:2015gna}:
\begin{align}
S_D = H_u \left( [L \nu^c]_{00} \sigma_{00} +[L \nu^c]_{02} \sigma_{01} + [L \nu^c]_{20} \sigma_{10} \right)
\end{align}

A simpler implementation would be to arrange instead that the Dirac neutrino matrix is diagonal in the familon basis from $S_D \sim H_u [L \nu^c]$, and after the seesaw with Majorana neutrino matrix from $S_M \sim [[\nu^c \nu^c] \bar{\phi}_{123}]$, the effective neutrinos inherit the eigenvectors originating from the relevant VEVs aligned by the appropriate alignment $S_V \sim  [\varphi \bar{\phi}_{123}]$.\footnote{
Note that with $L$ as a triplet and $\nu^c$ as an anti-triplet, the familon in the $[[\nu^c \nu^c] \bar{\phi}_{123}]$ invariant is also an anti-triplet, so the specific alignment superpotential required can depend on the implementation.
}

It could also be that there is more than one type of seesaw involved and related with the deviations from TBM \cite{Sierra:2014hea}. Here the analysis is performed at the level of the effective neutrino terms $S_\nu \sim H_u^2 [\phi_{123} [L L]]$. The frameworks presented here deviate from TBM but often preserve one of the TBM eigenvectors, making an eigenvector based approach particularly useful to study the mixing \cite{Sierra:2013ypa}.

In the following, different types of frameworks are illustrated through simple invariants for presentational purposes. These simpler examples can be the starting point to build complete models, but it remains necessary to distinguish the familons, and to allow this to be done through an auxiliary symmetry may require modifying the simpler terms for example with additional familon insertions (this procedure is illustrated for one of the examples in Section \ref{AS}).

\subsubsection{Invariant frameworks \label{I}}

In terms of neutrino invariants, the simplest type of $\Delta(27)$ framework relies exclusively on cubic invariants specific to $\Delta(27)$. They will be referred to as Invariant frameworks. One example relies only on triplet familons $\phi_{1}$ and $\phi_{123}$:
\begin{align}
S_\nu = H_u^2 \left(i_{1} [\phi_{1} [L L]_I]_{00} + s_{1} [\phi_{1} [L L]_S]_{00} + i_{123} [\phi_{123} [L L]_I ]_{00} + s_{123} [ \phi_{123} [L L]_S]_{00} \right)
\label{I_1_123}
\end{align}
where the coefficients are explicitly shown to be associated with the entries in the respective mass matrix
\begin{align}
M_\nu \propto
2 a_{1}
\begin{pmatrix}
	i_{1} & 0 & 0 \\
	0 & 0 & s_{1} \\
	0 & s_{1} & 0
\end{pmatrix}
+2 c
\begin{pmatrix}
	i_{123} & s_{123} & s_{123} \\
	s_{123} & i_{123} & s_{123} \\
	s_{123} & s_{123} & i_{123}
\end{pmatrix}
\end{align}

In this simple example the two familons couple in exactly the same way to the fermions, which is not compatible with separating their alignments as described above. A complete model requires some modification of the terms e.g. replacing the simpler $\phi_{1} [L L]_{I,S}]_{00}$ terms with $\phi_{1} [L L]_{I,S}]_{00} \sigma_{00}$.
The matrix structure is determined by the contractions and the VEVs. In this simple framework it leads to TBM mixing in the limit $i_{1}=s_{1}$, and in general the corresponding mixing scheme can be denoted as TM3 as it preserves the third TBM eigenvector $(0,1,-1)$. Like TBM, TM3 is not viable due to $\theta_{13} \neq 0$. An alternative but less predictive Invariant framework introduces a deviation from TM3 by using a non-trivial singlet familon such as $\sigma_{01}$, with terms 
\begin{align}
S_\nu = H_u^2 \left( [\phi_{123} [L L]_I ]_{00} + [ \phi_{123} [L L]_S]_{00} + [\phi_1 [L L]_I]_{00} +  [\phi_1 [L L]_S]_{00} + [\phi_1 [L L]_I]_{02} \sigma_{01} +  [\phi_1 [L L]_S]_{02} \sigma_{01} \right)
\label{I_1_123_s}
\end{align}
where the coefficients corresponding to the 6 invariants were omitted. In order to make this a complete model one could replace $\phi_{1} [L L]_{I,S}]_{00}$ with $\phi_{1} [L L]_{I,S}]_{00} \sigma_{00}$, where $\sigma_{00}$ and $\sigma_{01}$ would then have the same charge under an auxiliary symmetry and allow the triplet familons to be distinguished.

Another simple Invariant framework uses only triplet familons $\phi_{123}$ and $\phi_{23}$:
\begin{align}
S_\nu = H_u^2 \left( i_{123} [\phi_{123} [L L]_I ]_{00} + s_{123} [ \phi_{123} [L L]_S]+ i_{23} \phi_{23} [L L]_I]_{00} + s_{23} [\phi_{23} [L L]_S]_{00} \right)
\label{I_123_23}
\end{align}
\begin{align}
M_\nu \propto
2 c
\begin{pmatrix}
	i_{123} & s_{123} & s_{123} \\
	s_{123} & i_{123} & s_{123} \\
	s_{123} & s_{123} & i_{123}
\end{pmatrix}
+2 b
\begin{pmatrix}
	0 & s_{23} & -s_{23} \\
	s_{23} & -i_{23} & 0 \\
	-s_{23} & 0 & i_{23}
\end{pmatrix}
\end{align}

Like the previous examples, in order to have a complete model some variation of the terms is required in order to allow the two triplet familons to be distinguished.
This simple framework has a few interesting limits. For $i_{23} = s_{23}$ it preserves the second TBM eigenvector $(1,1,1)$, which corresponds to the tri-maximal mixing scheme TM2. For $i_{23} = -2 s_{23}$ it preserves the first TBM eigenvector $(-2,1,1)$, which corresponds to the tri-maximal mixing scheme TM1. In general if the TBM rotation is applied:
\begin{align}
V_{TBM}^\dagger M_\nu V_{TBM} \propto
2 c
\begin{pmatrix}
	i_{123}-s_{123} & 0 & 0\\
	0 & i_{123} & 0 \\
	0 & 0 & i_{123}-s_{123}
\end{pmatrix}
+2 b
\begin{pmatrix}
	0 & 0 & \frac{i_{23}+ 2s_{23}}{\sqrt{3}} \\
	0 & 0 & \sqrt{\frac{2}{3}} (-i_{23}+ s_{23}) \\
	\frac{i_{23}+ 2s_{23}}{\sqrt{3}} & \sqrt{\frac{2}{3}} (-i_{23}+ s_{23}) & 0
\end{pmatrix}
\end{align}
leaving the familon basis shows the particular TM2 and TM1 limits of this framework more clearly.

\subsubsection{$SU(3)$ frameworks \label{S}}

Similarly to the strategy described in \cite{deMedeirosVarzielas:2005ax} and implemented in the grand unification $\Delta(27)$ model \cite{deMedeirosVarzielas:2006fc}, this type of framework relies on invariants with anti-triplet familons e.g. 
\begin{align}
[L \bar{\phi}_{123}]_{00} [L \bar{\phi}_{123}]_{00}
\end{align}
which are invariant under $SU(3)$. They will be referred to as $SU(3)$ frameworks.\footnote{As $\Delta(27)$ is only providing the VEVs in $SU(3)$ frameworks, they can be used for other subgroups of $SU(3)$ like $T_7$ or $\Delta(6n^2)$ \cite{Varzielas:2012ss}.}
Justifying the absence of the additional invariants allowed by $\Delta(27)$ like $[L \bar{\phi}_{123}]_{10} [L \bar{\phi}_{123}]_{20}$ requires underlying assumptions about the messenger sector acting as the UV completion of the family symmetry model \cite{Varzielas:2010mp, Varzielas:2012ai}. In this case a possibility would be that the messengers in the neutrino sector are exclusively $\Delta(27)$ trivial singlets.
An advantage of $SU(3)$ frameworks is that they can be rather predictive, with only one effective invariant for each set of familons.

In order to illustrate the mass terms, one example relies on anti-triplet familons $\bar{\phi}_{123}$ and $\bar{\phi}_{23}$:
\begin{align}
S_\nu = H_u^2 \left( [L \bar{\phi}_{123}]_{00} [L \bar{\phi}_{123}]_{00} + [L \bar{\phi}_{23}]_{00} [L \bar{\phi}_{23}]_{00} + [L \bar{\phi}_{123}]_{00} [L \bar{\phi}_{23}]_{00} \right)
\label{S_123_23}
\end{align}
\begin{align}
M_\nu \propto
2 \bar{c}^2
\begin{pmatrix}
	1 & 1 & 1 \\
	1 & 1 & 1 \\
	1 & 1 & 1
\end{pmatrix}
+2 \bar{b}^2
\begin{pmatrix}
	0 & 0 & 0 \\
	0 & 1 & -1 \\
	0 & -1 & 1
\end{pmatrix}
+a \bar{c} \bar{b}
\begin{pmatrix}
	0 & -1 & 1 \\
	-1 & -2 & 0 \\
	1 & 0 & 2
\end{pmatrix}
\end{align}
where the coefficients of two terms were absorbed into the VEVs and a third effective parameter is chosen as the coefficient $a$ controlling the last term. A complete model using the same anti-triplet familons would require additional insertion of familons in some of the terms.
With $a=0$, this type of model leads to TBM  \cite{deMedeirosVarzielas:2006fc} (see also \cite{deMedeirosVarzielas:2008en}). The last term deviates TBM preserving the $(2,-1,-1)$ eigenvector which makes this an $SU(3)$ framework for the TM1 mixing scheme. This can be confirmed by leaving the familon basis through the TBM rotation to the mass matrix:
\begin{align}
V_{TBM}^\dagger M_\nu V_{TBM} \propto
\bar{c}^2
\begin{pmatrix}
	0 & 0 & 0 \\
	0 & 6 & 0 \\
	0 & 0 & 0
\end{pmatrix}
+\bar{b}^2
\begin{pmatrix}
	0 & 0 & 0 \\
	0 & 0 & 0 \\
	0 & 0 & 4
\end{pmatrix}
+ a \bar{c} \bar{b}
\begin{pmatrix}
	0 & 0 & 0 \\
	0 & 0 & - \sqrt{6} \\
	0 & - \sqrt{6} & 0
\end{pmatrix}
\end{align}

The consequences in terms of mixing angles are the same for any TM1 models and can be found in \cite{Varzielas:2012pa} and references therein. In this case the effective parameter $a$ is fixed by the observed value of $\theta_{13}$, which consequently predicts the deviations of the other angles from the TBM values. With only two other effective parameters, this TM1 framework is particularly predictive. It is clear that the lightest neutrino is massless, and on closer inspection the squared mass differences $\Delta m_{a}^2$, $\Delta m_{s}^2$ are controlled mostly by $\bar{b}^2$ and $\bar{c}^2$ respectively, requiring a mild hierarchy in the VEVs.

Alternatives to deviate from TBM in $SU(3)$ frameworks include replacing the $[L \bar{\phi}_{123}]_{00} [L \bar{\phi}_{23}]_{00}$ invariant with either $[L \bar{\phi}_{123}]_{00} [L \bar{\phi}_{3}]_{00}$ or $[L \bar{\phi}_{23}]_{00} [L \bar{\phi}_{3}]_{00}$.

\subsubsection{Alignment frameworks \label{A}}

Although the Invariant frameworks are minimal in terms of messengers and the $SU(3)$ frameworks are minimal in terms of effective parameters,
as all the frameworks rely on familon VEVs it is also interesting to consider frameworks with minimal requirements in terms of alignment fields and familons. They will be referred to as Alignment frameworks.

An interesting possibility to dispense with some alignments is to use effective familons.
The contraction of triplet familons $\phi_{123}$ and $\phi_{1}$ into anti-triplets leads to effective anti-triplet familons
\begin{align}
&[\phi_{123} \phi_{123}]_{I,S} \sim \bar{\phi}_{123} \label{eff123} \\
&[\phi_{1} \phi_{1}]_{I} \sim \bar{\phi}_{1} \,, \quad [\phi_{1} \phi_{123}]_{I} \sim \bar{\phi}_{1}\\
&[\phi_{1} \phi_{123}]_{S} \sim \bar{\phi}_{+} \,, \quad [\phi_{1} \phi_{123}]_{A} \sim \bar{\phi}_{23}
\end{align}
where the anti-symmetric contraction is particularly interesting as it leads to $\bar{\phi}_{23}$.
Therefore, an Alignment framework could rely on the effective anti-triplet familons e.g. 
\begin{align}
[L [\phi_{1} \phi_{123}]_{A}]_{00} [L [\phi_{1} \phi_{123}]_{A}]_{00} \sim [L \bar{\phi}_{23}]_{00} [L \bar{\phi}_{23}]_{00}
\end{align}
but unfortunately, the orthogonal direction $\langle  \bar{\phi}_{+} \rangle \propto (0,1,1)$ arises from the symmetric contraction $[\phi_{1} \phi_{123}]_{S} \sim \bar{\phi}_{+}$ and in $\Delta(27)$ the $I,S,A$ contractions transform in the same way. The possibility to select only anti-symmetric contractions exists in $\Delta(6n^2)$ groups and was exploited in \cite{Varzielas:2012ss}. For completeness, $[\phi_{23} \phi_{23}]_{I} \sim \bar{\phi}_{+}$ also features this direction, whereas $[\phi_{23} \phi_{23}]_{S} \sim \bar{\phi}_1$.

A more promising strategy is to combine terms from the Invariant framework (triplet familons) with terms from the $SU(3)$ framework (anti-triplet familons) to obtain Alignment frameworks with $\bar{\phi}_{123}$ and $\phi_{23}$, or with $\phi_{123}$ and $\bar{\phi}_{23}$. These Alignment frameworks have simpler requirements in terms of VEV alignment (because orthogonality conditions are between triplet and anti-triplet).

Alignment frameworks with  $\bar{\phi}_{123}$ and $\phi_{23}$ could appear as
\begin{align}
S_\nu = H_u^2 \left(s_{123} [L \bar{\phi}_{123}]_{00} [L \bar{\phi}_{123}]_{00} + i_{23} [\phi_{23} [L L]_I]_{00} +  s_{23} [\phi_{23} [L L]_S]_{00} + a_{I,S} [L \bar{\phi}_{123}]_{00} [L [\phi_{23} \phi_{23}]_{I,S}]_{00}  \right)
\end{align}
where if the $[L \bar{\phi}_{123}]_{00} [L [\phi_{23} \phi_{23}]_{I,S}]_{00}$ terms are absent, leads to models similar to those of eq.(\ref{I_123_23}) with TM2 and TM1 as limits, and one less parameter as here the term $[L \bar{\phi}_{123}]_{00} [L \bar{\phi}_{123}]_{00}$ corresponds to $i_{123} = s_{123}$. With the terms governed by $a_{I,S}$ present, the mixing scheme depends on many parameters although the simpler TM2 and TM1 limits could still be obtained by additionally imposing $a_{I}=-2a_{S}$, which is naturally verified if both vanish due to a specific UV completion or auxiliary symmetry.

Alignment frameworks with  $\phi_{123}$ and $\bar{\phi}_{23}$ could appear as
\begin{align}
S_\nu = H_u^2 \left(i_{123} [\phi_{123} [L L]_I]_{00} + s_{123} [\phi_{123} [L L]_S]_{00} + s [L \bar{\phi}_{23}]_{00} [L \bar{\phi}_{23}]_{00} + a [L \bar{\phi}_{23}]_{00} [L [\phi_{123} \phi_{123}]_{I,S}]_{00}  \right)
\label{A_123_23}
\end{align}
which is a very interesting Alignment framework.
As $[L [\phi_{123} \phi_{123}]_{I,S}]_{00}$ give the same structure due to eq.(\ref{eff123}), only the sum of the two contributions is relevant and denoted through $a$. 
This framework leads to TM1 models similar to those of eq.(\ref{S_123_23}), but with 4 relevant parameters. In comparison with the similar $SU(3)$ framework (with only 3 parameters), it allows non-zero determinant for $M_{\nu}$ and therefore a mass for the lightest neutrino. $\theta_{13}$ is directly related with $a$ which governs the TM1 deviations from TBM.

Additionally, within this Alignment framework (and the $SU(3)$ framework of eq.(\ref{S_123_23})) one can naturally explain the hierarchy between neutrino mass eigenstates by having a mild hierarchy in the VEVs of the two familons, which then establishes a relationship between the size of $\theta_{13}$ and $\frac{\Delta m_{s}^2}{\Delta m_{a}^2}$.\footnote{Relations between $\theta_{13}$ and $\Delta m_{s}^2/\Delta m_{a}^2$ are particularly interesting in the context of unified models like \cite{Varzielas:2012ss},
where $\theta_{13} \sim \sqrt { \frac{\Delta m_{s}^2}{\Delta m_{a}^2}} \sim 0.15$ can be further related to the size of the Cabibbo angle and to the hierarchy in quark masses (see also \cite{deMedeirosVarzielas:2005ax}).}

\subsection{Adding auxiliary symmetries \label{AS}} 
 
A complete framework matches a set of familon alignments with a set of fermion mass terms, without generating terms that invalidate the framework.
Typically this is achieved by adding an auxiliary symmetry which eliminates those terms, possibly in conjunction with specific UV completions \cite{Varzielas:2010mp, Varzielas:2012ai}.

The complete framework proposed here results from combining an auxiliary symmetry with a variation of the Alignment framework in eq.(\ref{A_123_23}) and a variation of the charged lepton terms in eq.(\ref{cl}). The effective familons strategy is employed requiring a minimal set of familons, $\phi_{3}$, $\sigma_{01}$, $\phi_{123}$ and $\bar{\phi}_{23}$:
\begin{align}
S_C &=  H_d \left( [L [\phi_3 \phi_3]_{I}]_{00} \tau^c + [L  [\phi_3 \phi_3]_{I}]_{02} \sigma_{01} \mu^c +  [L  [\phi_3 \phi_3]_{I}]_{01} \sigma_{01}^2 e^c \right)
\label{SC} \\
S_\nu &= H_u^2 (i_{123} [\phi_{123} [L L]_I]_{00} [\phi_{123} [\phi_{123} \phi_{123}]_{I,S}]_{00}  + s_{123} [\phi_{123} [L L]_S]_{00} [\phi_{123} [\phi_{123} \phi_{123}]_{I,S}]_{00} \label{Snu1} \\
&+[L [\phi_{123} \phi_{123}]_{I,S}]_{00} [L [\phi_{123} \phi_{123}]_{I,S}]_{00} + s [L \bar{\phi}_{23}]_{00} [L \bar{\phi}_{23}]_{00} + a [L \bar{\phi}_{23}]_{00} [L [\phi_{123} \phi_{123}]_{I,S}]_{00}  )
\label{Snu2}
\end{align}
where the $[L [\phi_{123} \phi_{123}]_{I,S}]_{00} [L [\phi_{123} \phi_{123}]_{I,S}]_{00}$ term acts like an effective $[L \bar{\phi}_{123}]_{00} [L \bar{\phi}_{123}]_{00}$, whose effect is absorbed by a suitable redefinition of the $i_{123}$ and $s_{123}$ couplings, and the $[\phi_{123} [\phi_{123} \phi_{123}]_{I,S}]_{00}$ contraction only affects the overall magnitude of the terms where it appears.

In the alignment sector, the minimal field content consists of alignment fields $\varsigma_{01}$,  $\varsigma_{11}$, $\varsigma_{21}$ (the set $S_{i1}$ resulting in eqs.(\ref{alt1},\ref{alt2},\ref{alt3}) discussed in Appendix \ref{align}) together with $\bar{\varphi}_{123}$ and a $\varsigma_{02}$:
\begin{align}
S_V &= a_{00} [\phi_{123} \bar{\varphi}_{123}]_{00} \sigma_{00} + a_{01} [\phi_{123} \bar{\varphi}_{123}]_{02} \sigma_{01} +\varsigma_{i1}[\phi_{3} \bar{\phi}_{23}]_{(-i)(2)} + \varsigma_{02} [\phi_{123} \bar{\phi}_{23}]_{01} \label{SV}
\end{align}

Although a specific UV completion will not be considered in full detail here, this framework relies on some underlying assumptions regarding the messengers. As mentioned in Section \ref{S}, neutrino messengers can avoid terms like $[L \bar{\phi}_{23}]_{01} [L \bar{\phi}_{23}]_{02}$. The absence of  $[L  [\phi_3 \phi_3]_{I}]_{00} \sigma_{00} \mu^c$, $[L  [\phi_3 \phi_3]_{I}]_{00} \sigma_{00}^2 e^c$ can be due to specific charged lepton messengers, which could be in this case non-trivial $\Delta(27)$ $1_{0i}$ singlets. The charged lepton messengers are distinct from neutrino messengers due to SM hypercharge.

The presence of the $\phi_3$ familon in $S_\nu$ or conversely the presence of $\phi_{123}$, $\bar{\phi}_{23}$ in $S_C$ would invalidate the framework, and likewise for terms in $S_V$. The familons need to be separated to avoid this.

Table \ref{fields} lists the field content together with symmetries and assignments for the set $S_{C}$ in eq.(\ref{SC}), $S_\nu$ in eqs.(\ref{Snu1},\ref{Snu2}) and $S_V$ in eq.(\ref{SV}), including an auxiliary $U(1)_{a}$ that eliminates terms that would invalidate the framework. The charges of alignment fields and fermions are expressed in terms of the familon charges, which are denoted by curly brackets (e.g. $\{ \phi_{3} \}$ is the $U(1)_{a}$ charge of $\phi_{3}$). Specific models correspond to a choice of the familon charges, and the existence of choices with $U(1)_{a}$ integer charges was explicitly verified: 2 (equivalent) choices remain for familon triplet charges $\{ \phi_{3} \} = \{\sigma_{00} \} = - \{ \phi_{123} \} = \pm 1$, 20 choices for integer charges ranging between $-2$ and $+2$ and many more for charges between $-3$ and $+3$. For each viable choice of charges it is possible to replace the continuous $U(1)_{a}$ symmetry with a sufficiently large cyclic subgroup $C_{n}$ without invalidating the model. Similarly the $R$-symmetry can be discrete \cite{Lee:2010gv, Lee:2011dya}.
Table \ref{fields} corresponds to a subset of models where $H_u$, $H_d$ are neutral under $U(1)_{a}$, for the sake of simplicity.

\begin{table}[tbp] \centering%
\begin{tabular}{|c||c||c||c||c|}
\hline
Field & $\Delta(27)$ & $SU(2)_{L}$ & $R$ & $U(1)_{a}$ \\ \hline\hline
$L$ & $\mathbf{3}$ & $\mathbf{2}$ & $\mathbf{1}$ & $-2 \{ \phi_{123} \}$ \\
$\tau^c$ & $1_{00}$ & $\mathbf{1}$ & $\mathbf{1}$ & $-2 \{ \phi_{3} \} + 2 \{ \phi_{123} \}$ \\
$\mu^c$ & $1_{00}$ & $\mathbf{1}$ & $\mathbf{1}$& $-2 \{ \phi_{3} \} + 2 \{ \phi_{123} \} - \{ \sigma_{00} \}$ \\
$e^c$ & $1_{00}$ & $\mathbf{1}$& $\mathbf{1}$ & $-2 \{ \phi_{3} \} + 2 \{ \phi_{123} \} - 2 \{ \sigma_{00} \}$ \\ \hline
$H_{u}$ & $1_{00}$ & $\mathbf{2}$ & $\mathbf{0}$ & 0 \\
$H_{d}$ & $1_{00}$ & $\mathbf{2}$ & $\mathbf{0}$ & 0 \\ \hline\hline
$\phi_{3}$ & $\mathbf{3}$ & $\mathbf{1}$ & $\mathbf{0}$ & $\{ \phi_{3} \}$ \\
$\phi_{123}$ & $\mathbf{3}$ & $\mathbf{1}$ & $\mathbf{0}$ & $\{ \phi_{123} \}$ \\
$\bar{\phi}_{23}$ &  $\mathbf{\bar{\mathbf{3}}}$ & $\mathbf{1}$ & $\mathbf{0}$ & $2 \{ \phi_{123} \}$ \\
$\sigma_{00}$ & $1_{00}$ & $\mathbf{1}$ & $\mathbf{0}$ & $\{ \sigma_{00} \}$ \\
$\sigma_{01}$ & $1_{01}$ & $\mathbf{1}$ & $\mathbf{0}$ & $\{ \sigma_{00} \}$ \\ \hline
$\bar{\varphi}_{123}$ &  $\mathbf{\bar{\mathbf{3}}}$ & $\mathbf{1}$ & $\mathbf{2}$ & $-\{ \phi_{123} \} - \{ \sigma_{00} \}$ \\
$\varsigma_{02}$ & $1_{02}$ & $\mathbf{1}$ & $\mathbf{2}$ & $-3 \{ \phi_{123} \}$  \\
$\varsigma_{i1}$ & $1_{i1}$ & $\mathbf{1}$ & $\mathbf{2}$ &  $-\{ \phi_{3} \} - 2 \{ \phi_{123} \}$ \\
 \hline
\end{tabular}%
\caption{Symmetries and Charges}\label{fields}%
\end{table}%

\section{Conclusion \label{sec:Conclusion}} 

In this paper $\Delta(27)$ is studied as a promising candidate for a family symmetry. The group has triplet and anti-triplet representations which makes it particularly suitable for grand unification, and has interesting CP properties.

Different options can provide the vacuum alignment of multiple family symmetry breaking familons. In supersymmetric frameworks there is D-term and F-term alignment. The latter was explored in Section \ref{sec:VEVs} to obtain vacuum alignment in directions $(0,0,1)$, $(0,-1,1)$, $(1,1,1)$ and also $(\omega,1,1)$.

Many frameworks for obtaining neutrino mixing were suggested in Section \ref{sec:masses}, including a simple predictive framework with only 3 parameters controlling directly the squared mass differences $\Delta m_{a}^2$, $\Delta m_{s}^2$ and $\theta_{13}$.

Viable frameworks can be constructed by combining a set of alignment terms and mass terms with an auxiliary symmetry. A minimal complete framework was presented in Section \ref{AS}.
  
\section*{Acknowledgements}

This project is supported by the European Union's Seventh Framework Programme for research, technological development and demonstration under grant agreement no PIEF-GA-2012-327195 SIFT.

\appendix

\section{$\Delta(27)$ \label{D27}}

$\Delta(27)$ has generators $c$ (for cyclic) and $d$ (for diagonal) with $c^3=d^3=1$.

The irreducible representations are 9 singlets and 2 triplets. The singlets $1_{ij}$ have $c_{1_{ij}}=\omega^i$ and $d_{1_{ij}}=\omega^j$, where $\omega \equiv e^{\mathrm{i} 2 \pi/3}$.
The two triplets can be denoted $3_{01}$ and $3_{02}$. The generator $c$ is represented equally for both and $d$ is represented as a diagonal matrix with entries that are powers of $\omega$ related to the subscripts of the triplet representation:
\begin{align}
c_{3_{ij}}=
\begin{pmatrix}
	0 & 1 & 0 \\
	0 & 0 & 1 \\
	1 & 0 & 0
\end{pmatrix}
\,, \quad
d_{3_{ij}}=
\begin{pmatrix}
	\omega^i & 0 & 0 \\
	0 & \omega^j & 0 \\
	0 & 0 & \omega^{-i-j}
\end{pmatrix}
\end{align}
$3_{01}$ and $3_{02}$ act as triplet and anti-triplet and are referred to as $\mathbf{3}$ and $\mathbf{\bar{3}}$ outside this Appendix. Singlets are obtained from $3_{01} \otimes 3_{02} = \sum_{i,j} 1_{ij}$. 2 triplets result in 3 anti-triplets and vice-versa: $3_{01} \otimes 3_{01} = [3_{02}]_{I} + [3_{02}]_{S}+[3_{02}]_{A}$, $3_{02} \otimes 3_{02} = [3_{01}]_{I} + [3_{01}]_{S}+[3_{01}]_{A}$.
Taking $A=(a_1,a_2,a_3)_{01}$ transforming as triplet $3_{01}$ (with lower indices) and $\bar{B}=(\bar{b}^1,\bar{b}^2,\bar{b}^3)_{02}$ transforming as anti-triplet $3_{02}$ (with upper indices),
the trivial singlet is 
\begin{align}
[A \bar{B}]_{00} \equiv (a_1 \bar{b}^1 + a_2 \bar{b}^2 + a_3 \bar{b}^3)_{00}
\label{AB00}
\end{align}
i.e. the $SU(3)$ invariant contraction. The non-trivial singlets can be built as
\begin{align}
[A \bar{B}]_{01} &\equiv (a_1 \bar{b}^3 + a_2 \bar{b}^1 + a_3 \bar{b}^2)_{01} \label{AB01} \\
[A \bar{B}]_{02} &\equiv (a_1 \bar{b}^2 + a_2 \bar{b}^3 + a_3 \bar{b}^1)_{02} \label{AB02} \\
[A \bar{B}]_{10} &\equiv (a_1 \bar{b}^1 + \omega^2 a_2 \bar{b}^2 + \omega a_3 \bar{b}^3)_{10} \label{AB10} \\
[A \bar{B}]_{11} &\equiv (\omega a_1 \bar{b}^3 + a_2 \bar{b}^1 + \omega^2 a_3 \bar{b}^2)_{11} \label{AB11} \\
[A \bar{B}]_{12} &\equiv (\omega^2 a_1 \bar{b}^2 + \omega a_2 \bar{b}^3 + a_3 \bar{b}^1)_{12} \label{AB12} \\
[A \bar{B}]_{20} &\equiv (a_1 \bar{b}^1 + \omega a_2 \bar{b}^2 + \omega^2 a_3 \bar{b}^3)_{20} \label{AB20} \\
[A \bar{B}]_{21} &\equiv (\omega^2 a_1 \bar{b}^3 + a_2 \bar{b}^1 + \omega a_3 \bar{b}^2)_{21} \label{AB21} \\
[A \bar{B}]_{22} &\equiv (\omega a_1 \bar{b}^2 + \omega^2 a_2 \bar{b}^3 + a_3 \bar{b}^1)_{22} \label{AB22}
\end{align}

The $I$, $S$, $A$ rules are the same for triplets and anti-triplets. The combination $I$ involves only $a_i b_i$ or $\bar{a}^i \bar{b}^i$:
\begin{align}
[A B]_{I} &\equiv (a_1 b_1, a_2 b_2, a_3 b_3)_{02} \label{TTI} \\
[\bar{A} \bar{B}]_{I} &\equiv (\bar{a}^1 \bar{b}^1, \bar{a}^2 \bar{b}^2, \bar{a}^3 \bar{b}^3)_{01} \label{AAI}
\end{align}

The symmetric $S$ and anti-symmetric $A$ combinations are:
\begin{align}
[A B]_{S} &\equiv (a_2 b_3 + a_3 b_2, a_3 b_1 + a_1 b_3, a_1 b_2 + a_2 b_1) \label{TTS}_{02} \\
[\bar{A} \bar{B}]_{S} &\equiv (\bar{a}^2 \bar{b}^3 + \bar{a}^3 \bar{b}^2, \bar{a}^3 \bar{b}^1 + \bar{a}^1 \bar{b}^3, \bar{a}^1 \bar{b}^2 + \bar{a}^2 \bar{b}^1)_{01} \label{AAS} \\
[A B]_{A} &\equiv (a_2 b_3 - a_3 b_2, a_3 b_1 - a_1 b_3, a_1 b_2 - a_2 b_1)_{02} \label{TTA} \\
[\bar{A} \bar{B}]_{A} &\equiv (\bar{a}^2 \bar{b}^3 - \bar{a}^3 \bar{b}^2, \bar{a}^3 \bar{b}^1 - \bar{a}^1 \bar{b}^3, \bar{a}^1 \bar{b}^2 - \bar{a}^2 \bar{b}^1) _{01} \label{AAA}
\end{align}

The transformation properties of all combinations can be checked by acting on $A$, $\bar{B}$ with the generators.

More details about $\Delta(27)$ and other $\Delta(3n^2)$ groups can be found in \cite{Luhn:2007uq, Ludl:2009ft, Ishimori:2010au}.

\section{F-term alignments in $\Delta(27)$ and similar groups \label{align}}

To discuss alignment options in more detail, in this appendix triplet alignment fields are referred as $A$, anti-triplet alignment fields as $\bar{B}$, triplet familons are $\theta$, and anti-triplet familons are $\alpha$ (with no bar, but upper indices). Singlets have labels of their representation, $\varsigma_{ij}$ for alignment fields and $\sigma_{ij}$ for familons. The $\langle \rangle$ notation for VEVs is dropped such that e.g. $\bar{\phi}_{23}^{1}=0$ implicitly refers to $\langle \bar{\phi}_{23}^{1} \rangle=0$.

Some of the best alignment options were already introduced in Section \ref{sec:VEVs} and used in eq.(\ref{SV}) of Section \ref{AS}.
Proceeding in a systematic fashion, one can start with the simplest renormalisable superpotentials.

\subsection{Triplet alignment field with familon triplet and familon singlets}

In terms of alignment fields, the choice is $A$,  $\bar{B}$, or one of nine singlets $\varsigma_{ij}$.
The basic invariant for triplet familon $\theta$ is then
\begin{align}
[\theta \bar{B}]_{00}
\end{align}
which would simply force the VEV to vanish.
The other renormalisable invariants involving only one alignment field and one familon $\theta$ are
\begin{align}
a_{I} [A [\theta \theta]_{I}]_{00} + a_{S} [A [\theta \theta]_{S}]_{00}
\end{align}
giving
\begin{align}
a_{I} \theta_1 \theta_1 + 2 a_{S} \theta_2 \theta_3 &= 0 \\
a_{I} \theta_2 \theta_2 + 2 a_{S} \theta_3 \theta_1 &= 0 \\
a_{I} \theta_3 \theta_3 + 2 a_{S} \theta_1 \theta_2 &= 0 
\end{align}
which leads to non-vanishing VEVs only for a special relation between the arbitrary couplings $a_{I}$ and $a_{S}$. Somewhat similar relations without this issue are obtained by adding one singlet familon $\sigma_{ij}$
\begin{align}
a_{00} [\theta \bar{B}]_{00} + a_{ij} [\theta \bar{B}]_{(-i)(-j)} \sigma_{ij}
\label{Bsigma}
\end{align}
allowing VEV directions that depend on the representation of the singlet $\sigma_{ij}$ (cf. eq.(\ref{SV}) which employed this type of invariants).
As discussed in Section \ref{sec:VEVs}, for familons $\sigma_{i0}$ the possibilities include $\theta \propto (1,0,0)$ and similar VEVs (i.e. those related by action of $\Delta(27)$ group elements, like the cyclic permutations $(0,1,0)$ and $(0,0,1)$).
For familons $\sigma_{0i}$ the possibilities include $\theta \propto (1,1,1)$, $(1,\omega,\omega^2)$ and similar VEVs.
For familons $\sigma_{ij}$ with $i,j \neq 0$ the possibilities include $\theta \propto (\omega,1,1)$ and similar VEVs. Although this last class of VEVs was not used in this paper, it is particularly relevant due to spontaneous geometrical CP violation \cite{Branco:1983tn,deMedeirosVarzielas:2011zw}, and had not been obtained previously in SUSY frameworks.

\subsection{Triplet alignment fields with familon triplet and anti-triplet}

If an anti-triplet $\alpha$ is present together with the triplet $\theta$, it can contribute to both the $A$ and $\bar{B}$ terms:
\begin{align}
a_{I} [A [\theta \theta]_{I}]_{00} + a_{S} [A [\theta \theta]_{S}]_{00} + a [A \alpha]_{00} \\
b_{I} [[\alpha \alpha]_{I} \bar{B}]_{00} + b_{S} [[\alpha \alpha]_{S} \bar{B}]_{00} + b [\theta \bar{B}]_{00}
\end{align}
the F-terms with respect to the alignment field triplet components $A_i$ would then give
\begin{align}
a_{I} \theta_1 \theta_1 + 2 a_{S} \theta_2 \theta_3 + a \alpha^1 &= 0 \\
a_{I} \theta_2 \theta_2 + 2 a_{S} \theta_3 \theta_1 + a \alpha^2 &= 0 \\
a_{I} \theta_3 \theta_3 + 2 a_{S} \theta_1 \theta_2 + a \alpha^3 &= 0
\end{align}
which can relate the alignment between an anti-triplet familon $\alpha$ and triplet familon $\theta$, but is not sufficient to constrain the direction of either. Nevertheless, if one of the familons is separately aligned in a direction in the class $(1,0,0)$ or $(1,1,1)$, that special direction is passed into the other familon through this type of term, but this doesn't apply to other directions.
Similarly from the F-terms with respect to the alignment field anti-triplet components $\bar{B}^i$
\begin{align}
b_{I} \alpha^1 \alpha^1 + 2 b_{S} \alpha^2 \alpha^3 + b \theta_1 &= 0 \\
b_{I} \alpha^2 \alpha^2 + 2 b_{S} \alpha^3 \alpha^1 + b \theta_2 &= 0 \\
b_{I} \alpha^3 \alpha^3 + 2 b_{S} \alpha^1 \alpha^2 + b \theta_3 &= 0
\end{align}

As the directions passed between familons only remain invariant for special directions, combining the triplet alignment field and the anti-triplet alignment field with arbitrary parameters should only allow special solutions. In addition this fixes the absolute magnitude of both VEVs. A simple example of this occurs for the solution where both familons mutually align in the $(1,0,0)$ direction:
\begin{align}
\theta_1 = - \frac{b_{I}}{b} (\alpha^1)^2 \,, \quad \alpha^1 = - \frac{a_{I}}{a} (\theta_1)^2 \\
(\theta_1)^3=-\frac{b a^2}{b_{I} a_{I}^2} \,, \quad (\alpha^1)^3=-\frac{a b^2}{a_{I} b_{I}^2}
\end{align}
and similarly for the $(1,1,1)$ direction where the symmetric coefficients are involved e.g. $\theta_1 = \theta_2 = \theta_3 = - \frac{b_{I}+2 b_{S}}{b} (\alpha_1)^2$.

\subsection{Singlet alignment fields with familon triplet and anti-triplet}

Without an anti-triplet familon to couple to, there is no renormalisable coupling of $\theta$ to any singlet alignment field $\varsigma_{ij}$.
But with an anti-triplet familon $\alpha$:
\begin{align}
\varsigma_{ij} [\theta \alpha]_{(-i)(-j)}
\end{align}
which enforces a relation between the triplet and anti-triplet components which is a kind of singlet specific orthogonality condition between $\theta$, $\alpha$ (cf. eq.(\ref{SV}) which employed this type of invariants).

If there are multiple alignment field singlets it is possible to restrict the possible directions. One example is
\begin{align}
S_{i0}= a_{00} \varsigma_{00} [\theta \alpha]_{00} + a_{10} \varsigma_{10} [\theta \alpha]_{20} +  a_{20} \varsigma_{20} [\theta \alpha]_{10}
\end{align}
where the label of the coefficients corresponds to the alignment field singlet. The F-terms give
\begin{align}
a_{00} (\theta_1 \alpha^1 + \theta_2 \alpha^2 + \theta_3 \alpha^3) &= 0 \\
a_{10} (\theta_1 \alpha^1 + \omega \theta_2 \alpha^2 + \omega^2 \theta_3 \alpha^3) &= 0 \\
a_{20} (\theta_1 \alpha^1 + \omega^2 \theta_2 \alpha^2 + \omega \theta_3 \alpha^3) &= 0 
\end{align}
and summing the 3 equations leads to 
\begin{align}
(\theta_1 \alpha^1) &= 0 
\end{align}
but one can also sum the 3 while multiplying specific powers of $\omega$ to isolate:
\begin{align}
(\theta_2 \alpha^2) &= 0 \\
(\theta_3 \alpha^3) &= 0 
\end{align}
meaning this set of 3 alignment fields enforces one of the two familons to have two vanishing entries, while the other familon must have the other one vanishing. This is a very interesting option to simultaneously obtain a $\phi_{1}$ familon with $\phi_{1_2}= \phi_{1_3}= 0$ while guaranteeing $\bar{\phi}_{23}^{1}=0$.

Similarly, $\varsigma_{01}$, $\varsigma_{11}$, $\varsigma_{21}$ (note the second label is the same on all three, as in $S_{i0}$)  constituting $S_{i1}$ would lead to
\begin{align}
(\theta_3 \alpha^1) &= 0 \label{alt1}\\
(\theta_1 \alpha^2) &= 0 \label{alt2}\\
(\theta_2 \alpha^3) &= 0 \label{alt3}
\end{align}
which is particulary interesting for a $\phi_{3_3}\neq 0$, $\bar{\phi}_{23}^{1}=0$ solution.
The set $\varsigma_{02}$, $\varsigma_{12}$, $\varsigma_{22}$ constituting $S_{i2}$ leads to
\begin{align}
(\theta_2 \alpha^1) &= 0 \\
(\theta_3 \alpha^2) &= 0 \\
(\theta_1 \alpha^3) &= 0 
\end{align}

Combining two of these 3 singlet sets (a total of 6 singlet alignment fields for the same pair of triplet and anti-triplet familons) restricts the directions such that both familons have 2 zero entries: the same non-zero entry for $S_{i1} + S_{i2}$ and either the cyclic pairs for 
$S_{i0} + S_{i2}$ (from triplet to anti-triplet, e.g. $\theta_{1}\neq0$ together with $\alpha^{2}\neq0$)
or the anti-cyclic pairs for
$S_{i0} + S_{i1}$ (from triplet to anti-triplet, e.g. $\theta_{1}\neq0$ together with $\alpha^{3}\neq0$).
Adding another singlet alignment field to one of these sets of 6 makes one of the familon VEVs vanish.

Other sets of alignment field singlets include:
\begin{align}
S_{0i}= a_{00} \varsigma_{00} [\theta \alpha]_{00} + a_{01} \varsigma_{01} [\theta \alpha]_{02} +  a_{02} \varsigma_{02} [\theta \alpha]_{01}
\end{align}
giving
\begin{align}
a_{00} (\theta_1 \alpha^1 + \theta_2 \alpha^2 + \theta_3 \alpha^3) &= 0 \\
a_{01} (\theta_3 \alpha^1 + \theta_1 \alpha^2 + \theta_2 \alpha^3) &= 0 \\
a_{02} (\theta_2 \alpha^1 + \theta_3 \alpha^2 + \theta_1 \alpha^3) &= 0 
\end{align}
from which one can sum the 3 to obtain
\begin{align}
(\theta_1 + \theta_2 + \theta_3) (\alpha^1 + \alpha^2 + \alpha^3)&= 0 
\end{align}
i.e. either one of the sums or both sums vanish in the complex plane. One can replace one of these solutions in the original equations, or
sum the 3 equations with appropriate powers of $\omega$ to obtain equivalently 
\begin{align}
(\theta_1 + \omega \theta_2 + \omega^2 \theta_3) (\alpha^1 + \omega^2 \alpha^2 + \omega \alpha^3)&= 0 \\
(\theta_1 + \omega^2 \theta_2 + \omega \theta_3) (\alpha^1 + \omega \alpha^2 + \omega^2 \alpha^3)&= 0
\end{align}
which makes it more evident that if the triplet components obey $(\theta_1 + \theta_2 + \theta_3)=0$, then the anti-triplet components obey both $(\alpha^1 + \omega \alpha^2 + \omega^2 \alpha^3)=0$ and $(\alpha^1 + \omega^2 \alpha^2 + \omega \alpha^3)=0$, for example if the triplet VEV is $(1,\omega,\omega^2)$ the respective anti-triplet VEV is $(1,1,1)$ (and vice-versa).
Similarly, a different set of 3 alignment field singlets $\varsigma_{10}$, $\varsigma_{11}$, $\varsigma_{12}$ (note the first label is the same on all three, as in $S_{0i}$) constituting $S_{1i}$ would lead to 
\begin{align}
(\theta_1 + \theta_2 + \theta_3) (\alpha^1 + \omega \alpha^2 + \omega^2 \alpha^3)&= 0 \\
(\theta_1 + \omega \theta_2 + \omega^2 \theta_3) (\alpha^1 + \alpha^2 + \alpha^3)&=0 \\
(\theta_1 + \omega^2 \theta_2 + \omega \theta_3) (\alpha^1 + \omega^2 \alpha^2 + \omega \alpha^3)&=0
\end{align}
whereas the set of 3 alignment fields $\varsigma_{20}$, $\varsigma_{21}$, $\varsigma_{22}$ that would constitute $S_{2i}$ would lead to 
\begin{align}
(\theta_1 + \theta_2 + \theta_3) (\alpha^1 + \omega^2 \alpha^2 + \omega \alpha^3)&= 0 \\
(\theta_1 + \omega^2 \theta_2 + \omega \theta_3) (\alpha^1 + \alpha^2 + \alpha^3)&= 0 \\
(\theta_1 + \omega \theta_2 + \omega^2 \theta_3) (\alpha^1 + \omega \alpha^2 + \omega^2 \alpha^3)&= 0
\end{align}
and combining 6 alignment field singlets narrows down the solutions. Among the 3 remaining solutions with both familons non-vanishing for $S_{1i}+S_{2i}$ is the $(1,1,1)$ VEV for both triplet and anti-triplet.

\subsection*{Summary and applications for other groups}

In order to align triplet or anti-triplet familon VEVs with renormalisable superpotential terms in $\Delta(27)$, one must necessarily have another anti-triplet or triplet field and there are three possibilities. The first is the anti-triplet (or triplet) is an alignment field and one can obtain relevant VEVs in conjuction with additional familons singlets. The second is both the alignment fields and the additional familons are triplets. The third option is having singlet alignment fields, and one can obtain relevant VEVs in conjuction with triplet and anti-triplet familons.
The main results are summarised in Table \ref{av} (where triplet and anti-triplets can be reversed).

\begin{table}[h] \centering%
\begin{tabular}{|c|c|}
\hline
Alignment fields & Familon VEVs \\ \hline\hline
Anti-triplet & $1_{i0}$,$1_{j0}$; Triplet $(1,0,0)$ class \\
Anti-triplet & $1_{0i}$,$1_{0j}$; Triplet $(1,1,1)$ class \\
Anti-triplet & $1_{ij}$,$1_{kl}$; Triplet $(\omega,1,1)$ class \\ \hline
Anti-triplet and triplet & Triplet and anti-triplet $(1,0,0)$ class\\
Anti-triplet and triplet & Triplet and anti-triplet $(1,1,1)$ class \\ \hline
3 Singlets $1_{i0}$ & Triplet $(1,0,0)$ and anti-triplet $(0,y,z)$ \\ 
3 Singlets $1_{0i}$ & Triplet $(1,1,1)$ and anti-triplet $(1,\omega,\omega^2)$ \\
 \hline
\end{tabular}%
\caption{Alignments in $\Delta(27)$}\label{av}
\end{table}%

The D-term alignments found in $\Delta(27)$ could be used in other groups, and the same is true for the F-term alignments.
Given that the product rules for $T_7$ triplet, anti-triplet and singlets are so similar to those of $\Delta(27)$, many of the F-term alignments discussed in this Appendix can be directly applied to $T_7$ frameworks - namely, the options that do not involve $\Delta(27)$ singlets other than the three $1_{i0}$. This includes some of the options in eq.(\ref{Bsigma}) leading to $(1,0,0)$, the mutual alignment option which relies only on pairs of triplet and anti-triplet leading to both being aligned as $(1,0,0)$ or both being aligned as $(1,1,1)$, and the $S_{i0}$ option which relies on 3 alignment field singlets to align a pair of triplet and anti-triplet where one of them has two zeros and the other is orthogonal, e.g. $(1,0,0)$ and $(0,y,z)$.

The $S_{i0}$ option is particularly versatile as it does not rely on the product of two triplets or on the product of two anti-triplets, and as such can be used for $\Delta(3 n^2)$ and $\Sigma(3n^3)$ groups in general.
$\Delta(3n^2)$ and $\Sigma(3n^3)$ groups with $n$ multiple of 3 (e.g. $\Sigma(81)$) have 9 singlets like $\Delta(27)$. For such groups all options in eq.(\ref{Bsigma}) and those involving sets of alignment field singlets beyond $S_{i0}$ are available to align all of the directions discussed here.

\pagebreak

\end{document}